\documentclass[twoside,onecolumn]{article}

\usepackage[title]{appendix}

\usepackage{blindtext} % Package to generate dummy text throughout this template 

\usepackage[english]{babel} % Language hyphenation and typographical rules

\usepackage[hmarginratio=1:1,top=32mm,columnsep=20pt]{geometry} % Document margins
\usepackage[hang, small,labelfont=bf,up,textfont=it,up]{caption} % Custom captions under/above floats in tables or figures
\usepackage{booktabs} % Horizontal rules in tables

\usepackage{lettrine} % The lettrine is the first enlarged letter at the beginning of the text

\usepackage{enumitem} % Customized lists
\setlist[itemize]{noitemsep} % Make itemize lists more compact

\usepackage{abstract} % Allows abstract customization
\usepackage{titlesec} % Allows customization of titles
\titleformat{\section}[block]{\large\scshape\centering}{\thesection.}{1em}{} % Change the look of the section titles
\titleformat{\subsection}[block]{\large}{\thesubsection.}{1em}{} % Change the look of the section titles

\usepackage{fancyhdr} % Headers and footers
\usepackage{titling} % Customizing the title section
\usepackage{hyperref} % For hyperlinks in the PDF

\usepackage{lineno}
\usepackage{amsmath,amsthm,amssymb}
\usepackage{afterpage}
\usepackage{array}
\usepackage{booktabs}
\usepackage{caption}
\usepackage{geometry}
\usepackage{footnote}
\usepackage{multirow}
\usepackage{threeparttable,rotating}
\usepackage{pdflscape}
\usepackage{setspace}
\usepackage{subcaption}
\usepackage{tabularx}
\usepackage{url}
\usepackage{bibentry}
\usepackage[numbers]{natbib}
\usepackage{eurosym}
\usepackage{graphicx}
\usepackage{float}
\usepackage{notoccite}
\usepackage{tabu}
\usepackage{mwe}

\def\beq{\begin{equation}}
\def\eeq{\end{equation}}
\def\ie{\textit{i.e.}}
\def\etal{\textit{et al.}}

\pretitle{\begin{center}\Huge\bfseries} % Article title formatting
\posttitle{\end{center}} % Article title closing formatting
\title{The Network Effect in Credit Concentration Risk} % Article title
\author{%
\textsc{Davide Cellai} 
\thanks{The views expressed in the paper are those of the authors and do not represent the views of the
 Central Bank of Ireland or the European Central Bank/Single Supervisory Mechanism.} 
 \\[1ex] % Your name
\normalsize RADAR, Central Bank of Ireland, North Wall Quay, Dublin 1, Ireland \\ % Your institution
\normalsize Department of Mathematics and Statistics, University of Limerick, Ireland \\ % Your institution
\normalsize \href{mailto:davide.cellai@centralbank.ie}{davide.cellai@centralbank.ie} % Your email address
\and % Uncomment if 2 authors are required, duplicate these 4 lines if more
\textsc{Trevor Fitzpatrick {\small*}} 
 \\[1ex] % Second author's name
\normalsize RADAR, Central Bank of Ireland, North Wall Quay, Dublin 1, Ireland \\ 
\normalsize Southampton Business School, University of Southampton, Highfield, Southampton, SO17 1BJ, UK \\ 
\normalsize \href{mailto:trevor.fitzpatrick@centralbank.ie}{trevor.fitzpatrick@centralbank.ie} % Second author's email address
}
\date{\today} % Leave empty to omit a date
\renewcommand{\maketitlehookd}{%
\begin{abstract}
\noindent 
Measurement and management of credit concentration risk is critical for banks and relevant for micro-prudential requirements. While several methods exist for measuring credit concentration risk within institutions, the systemic effect of different institutions' exposures to the same counterparties has been less explored so far. 
In this paper, we propose a measure of the systemic credit concentration risk that arises because of common exposures between different institutions within a financial system. This approach is based on a network model that describes the effect of overlapping portfolios. 
This metric is applied to synthetic and real world data to illustrate that the effect of common exposures is not fully reflected in single portfolio concentration measures.
It also allows to quantify several aspects of the interplay between interconnectedness and credit risk. % <- this sentence could be removed if needed
Using this network measure, we formulate an analytical approximation for the additional capital requirement corresponding to the systemic risk arising from credit concentration interconnectedness.
Our methodology also avoids double counting between the granularity adjustment and the common exposure adjustment. 
Although approximated, this common exposure adjustment is able to capture, with only two parameters, an aspect of systemic risk that can extend single portfolios view to a system-wide one.
\end{abstract}
}

\begin{document}
\maketitle

\section{Introduction}
\label{section introduction}

Concentration risk arises in a credit portfolio when there is an uneven distribution of exposures in the portfolio. This occurs in two ways: individual borrowers (single-name concentration) or groups of borrowers aggregated by sectors or geographical regions (sectoral concentration).

Concentrations are important because the theoretical foundations of Basel risk-based capital requirements capital requirements assume infinite granularity of a portfolio. The  are based on the Asymptotic Single Risk Factor (ASRF) model, assuming an infinitely large pool of loans resulting in a diversified credit risk portfolio \cite{Gordy:2003}. This assumption enables an analytical solution for capital requirement calculation.  Because of this, additional capital is required against this risk \cite{basel:2006bis}. 

From a micro-prudential perspective, credit concentration has been associated with increased risk of bank failures and increased magnitude of these failures when they occur. Credit concentrations are also important to macro-prudential supervisors when they have the responsibility of identifying and mitigating sources of systemic risk across the financial system.  Several banks lending to the same counterparty is an example of system risk arising from credit concentration.

The consequences of not appropriately measuring and mitigating credit concentration risk can be catastrophic. It has been a hallmark of several recent financial crises in Ireland and other OECD economies. Concentrations arise at several layers of disaggregation: first, at a sector level such as property, within that sector at a commercial property level, and then at a key number of large developers within that sector. Lending in this sector is particularly cyclical, and therefore subject to more abrupt declines in collateral values. 

In the Irish crisis and the subsequent parliamentary inquiry that followed, one senior executive of an Irish credit institution recounted that the top 30 exposures accounted for over 50\% of the bank's total exposure, and 48\% of its profit \footnote{Large Exposures/Cross-Bank Lending section of the Irish Banking Inquiry Report, Volume 1, \url{https://inquiries.oireachtas.ie/banking/volume-1-report/chapter-1/} .}. That firm was subsequently liquidated. This problem has occurred previously as several banks suffered large losses related to Worldcom, Enron, and Parmalat, as well as the large commercial real estate losses experienced by US banks in the late 1980s \cite{basel:2006bis}.

In practice, bank credit portfolios can have concentrated exposures relating to specific obligors or sectors depending on their business models. To capture this specific aspect of credit risk, credit concentration risk is specifically addressed in the Supervisory Review and Evaluation Process (SREP) as a capital surcharge, under Pillar 2.  In Pillar 2 capital requirements, concentration risk charges apply to an individual institution. However, individual institutions can have overlapping portfolios of obligors, some of which may be important to individual banks and to the system as a whole. In this paper, we focus on name concentration - typically a feature of Corporate /Commercial Real Estate or other types of specialised lending.

We make three main contributions to existing research. 
The first one conceptualises credit risk using a system-wide approach based on network science. We integrate two strands of the research literature, credit risk measurement and network science, providing for a metric that measures interdependence due to overlapping portfolios.
This approach is  general and could be applied to other types of assets, sectors or geographic areas. 
Second,   we demonstrate how this system wide credit concentration risk measure can be practically applied, by providing an analytical estimate of the additional prudential capital that corresponds to this type of credit risk concentration.  
Third, using simulated and real-world data, we show how this additional capital formula can be calibrated with respect to existing capital requirements and to avoid double counting with the granularity adjustment.

\section{Related work and research contribution}
\label{section related work research}

Various proposals exist for calculation of name or sector concentration risk to adjust the ASRF approach. For single-name concentration, existing methods are based on some form of granularity adjustment. These include the Herfindhal-Hirschmann Index or Gini. Depending on  the implementation, these measures may not reflect changes in obligor risk \cite{Gordy:2013, dullmann:2007, Cabedo:2011}.  
 
Sector concentration risk can be calculated using multi-factor models taking into account the systematic risk associated with that sector \cite{pykhtin:2004}, or derive analytical approximations that require simplifying assumptions such as intra-sector correlations remaining constant \cite{kurtz:2018}. 
% See Kurtz and Eva L¨utkebohmert paper that I just sent you; there's a couple of other approaches that would be usefully included here too. They may be referees! 

Since the 2007 crisis, various types of inter-disciplinary approaches to measuring concentration risk have been developed. 
The literature merging network theory with various approaches to measuring systemic risk includes work by Alter \etal \cite{AdrianAlter:2015ta}, and Levy-Carciente {\etal} \cite{LevyCarciente:2015cta}, as well as earlier work by Huang \etal \cite{Huang:2013fu} and Caccioli \etal \cite{Caccioli:2014bfa}. In their frameworks, similar to ours, they use bipartite networks to describe interconnectedness through various measures. 
However, the primary purpose of their analysis is to use the asset side of the bank balance sheet as part of mechanism of cascading failures driven by links through the interbank market. This is used to investigate the dynamics of cascading bank failures in response to an initial shock to asset value. 

In our approach, we are interested in identifying the aspects of structural systemic risk with a focus on borrowers as the source of risk, rather than the dynamics of a cascading interbank failure.  Our work focuses on distinguishing the effect of common exposures across banks, and not within banks. While this network effect is implicitly encapsulated in the works by Huang \etal \cite{Huang:2013fu} and Caccioli \etal \cite{Caccioli:2014bfa}, we wish to make it explicit,  and develop a measure that does not require long Monte Carlo simulations and can be used to calculate additional capital requirements specifically related to this type of risk. 
The paper by Das \cite{Das:2016ii} focusses on measuring systemic risk in a flexible manner with the help of network based properties. However, he does not develop how network metrics interact with the existing regulatory framework.
Our work has also some common elements with that by Alter \etal \cite{AdrianAlter:2015ta}, in that we link common exposures to across the system to capital requirements. However, our credit risk mechanism is very different to the Creditmetrics approach used in their work. 

%Given the focus on determining systemic risk, the issue of varying risk or connectivity of individual exposures within  the system or how this interacts with the existing regulatory framework for banks is not fully developed in that paper. Therefore, our focus is on overlapping exposures and propose a methodology adopting network science tools that extend traditional measures of credit concentration to quantify the potential capital that could be required to address this risk. 
%
%We draw on network science, in particular the paper by Zhou et al \cite{Zhou:2007hx} combining network analysis to develop a recommendation system. For our purposes, this paper defines a similarity metric based on a portfolio of purchased products \footnote{In recommendation systems, a model typically involves a set of users and a set of products. This system can also be represented by a bipartite network, characterized by a layer of users and a layer of products.} 
%
%Each user is connected to all the products that she has bought.
%Zhou et al define a metric that describes the similarity of two users based on their portfolio of collected products. 
%Users' similarity can then  be exploited by recommending those products that may appeal to users with similar characteristics.
%The original metric is defined on an unweighted network, but the definition can be naturally extended to weighted networks \cite{Liu:2009}.

We draw on network science, in particular the paper by Zhou {\etal} \cite{Zhou:2007hx}, to build a projected network from a bipartite network and quantify the impact of lenders onto each others.
The work by Zhou {\etal} has been developed for a different purpose (strategies in recommendation systems), but  the logic of their model is also useful in addressing our problem.
%The original metric is defined on an unweighted network, but the definition can be naturally extended to weighted networks \cite{Liu:2009}.

\section{Methodology}
\label{section model}

Based on the existing work and the motivating examples given in the introduction, we have several requirements for a framework that can fully describe interconnectedness due to common exposures. 
First, the framework should take into account not only the presence of a connection, but its magnitude and risk. Second, it should be tractable enough to decompose individual exposure contributions and their effect overall credit concentration risk within a given system. Third, it should also be able to quantify the effect concentrations within one bank on other banks. Finally, the framework should calculate the effect of changes in risky exposures through changes in the network topology. As the focus is on credit concentration risk, we only consider a simplified subset of a complex system involving lenders and borrowers.

We consider a set of $n$ lenders and $m$ borrowers, with the following simplifications:
  \begin{enumerate}
    \item We assume that lenders and borrowers are two distinct sets (e.g. we neglect interbank loans).
    \item We neglect lenders liabilities structures.
  \end{enumerate}
These assumptions are reasonable as we want to focus on the structural nature of systemic risk, building on within single institution credit concentration risk.   

\begin{figure}[htb]
	\centering
  \includegraphics[width=0.9\columnwidth]{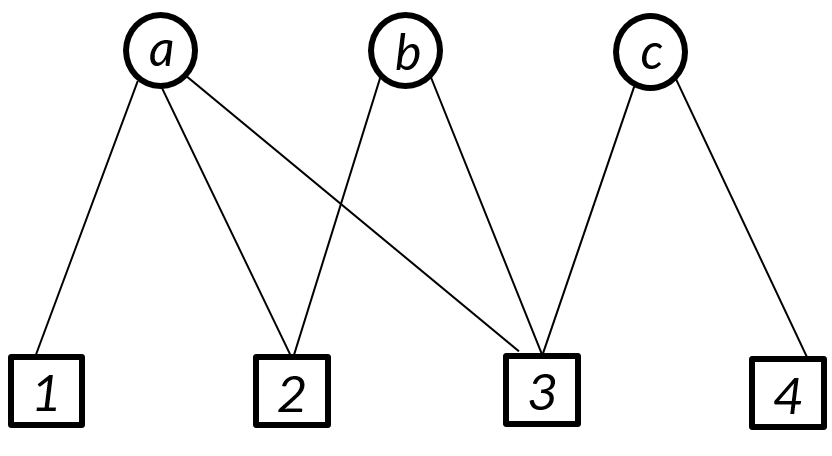}
  \caption{Representation of a system of $n=3$ lenders and $m=4$ borrowers as a bipartite network.}
  \label{fig:bipartite}
\end{figure}

This problem can be represented as a bipartite network where one layer corresponds to the lenders and the other to their borrowers (Fig.~\ref{fig:bipartite}).The link between a lender and a borrower represents the exposure of the lender to the debt of the considered borrower.

%
%

%\section{Systemic effect of co-exposures}

A common measure of portfolio concentration (e.g. for a bank) is the Herfindhal-Hirschman Index (HHI):
\begin{equation}
	HHI = \frac{\sum_k E_k^2}{\left(\sum_k E_k\right)^2} ,
	\label{eq:HHI}
\end{equation}
where the sums are calculated over the borrowers and $E_k$ is the exposure of the considered lender to the borrower $k$.

The HHI is a simple metric that can be used to measure the credit concentration risk at the level of each considered institution. 
However, pairs of lenders may be exposed to the same borrowers, generating further concentration risk on a system wide level. In practice, the HHI does not take into account the potential risk inherent to common exposures.

As mentioned earlier, our problem has similarities with the one investigated by Zhou et al \cite{Zhou:2007hx}.
In our case, we  have a bipartite network that comprises two layers: a layer of financial institutions (e.g. banks) and a layer of borrowers.
Similarly to the network of users and products, here we wish to quantify the similarities in the composition of financial institutions portfolios.
We therefore define a Dependency Index $DI$ based on a modification of the quantity proposed by \cite{Zhou:2007hx}. Our Dependency Index is defined as the metric in \cite{Zhou:2007hx} except for the fact that the considered bipartite network of banks and counterparties is weighted, as in \cite{Liu:2009}.

To calculate the Dependency index,  we rescale exposures in a way that takes into consideration the risk of the corresponding borrower.
Let $e_{ik}$ be the exposure of lender $i$ to the borrower $k$.
We define the risk adjusted exposure $w_{ik}$ of lender $i$ to borrower $k$ as 
\beq
	\label{eq:rw-exposures}
	w_{ik} = f(r_k) e_{ik}
\eeq
where $f(r_k)$ is a function of $r_k$, the estimated risk of entity $k$. This function modifies the unweighted exposure into a weighted exposure based on a risk measure for entity $k$. This risk measure could be a credit rating, and internal credit scale or probability of default (PD), or a market based metric that can rank order credit risk, such as credit spreads. 
As a result, $w_{ik}$ is a $n\times m$ matrix that connects $n$ lenders with $m$ borrowers.

At this stage, we project the bipartite network onto the lenders' layer.
In other words, we create a single-layer network where a directional link from lender $i$ to lender $j$ represents the size of impact of $i$ to $j$, corresponding to the systemic effect of credit concentration due to common exposures.
The adjacency matrix $S = (s_{ij})$ of this projection is calculated as
\begin{equation}
	s_{ij} = \sum_{l=1}^m \frac{w_{il}w_{jl}}{ \sum_{p=1}^n w_{pl} \sum_{q=1}^m w_{jq} },
	\label{eq:matrix-s-definition}
\end{equation}
where $w_{il}$ is the element of the adjacency matrix of the bipartite network (from lender $i$ to borrower $l$).
The sum over $l$ in Equation (\ref{eq:matrix-s-definition}) runs over the $m$ borrowers. As a result, we obtain a $n\times n$ matrix that only depends on the lenders.
We name $S$ \emph{impact matrix}, as it represents the impact that one portfolio has onto another one due to common exposures.
More specifically, given two distinct lenders $i$ and $j$, the element matrix $s_{ij}$ represents the impact that lender $i$ has on lender $j$. 
For example, if the two portfolios have a large overlap, $s_{ij}$ will be higher.
On the other hand, the matrix is not necessarily symmetric, i.e. $s_{ij}\neq s_{ji}$, whenever the two lenders have different total exposures.
It is easy to see that the impact of a large lender on a small one will be larger than the one of a small lender to a large one.
Finally, each element on the diagonal $s_{ii}$ represents a measure of the amount of exposures of lender $i$ that are not connected to other lenders. 

%This network is weighted, directed, and has self loops.
%As in \cite{Zhou:2007hx}, we can interpret this matrix in a visual way as it follows.
%Let us assume that each lender as a bucket containing an amount of fluid proportional to their total exposures. The borrowers are also buckets that are initially empty. The fluid is dyed with a color that uniquely identifies the lender where it is stored.
%Let us also assume that the links between lenders and borrowers are pipes of sizes that are proportional to each exposure. 
%Now, we consider a process made of two steps.
%First, let us imagine a flow of the fluid from the lenders to the borrowers.
%Each borrower will receive fluid from each lender in a way that is proportional to the weight of each link.
%In general, each bucket borrower will contain some of the fluid coming from different lenders (and we assume the colors do not mix).
%Second, let us now imagine that all the fluid that has flowed to the borrowers goes back to the lenders, each colored fluid going to the lenders in a  way that is proportional to the size of the links/pipes.
%At the end of this process, the amount of fluid of each color that has reached each lender corresponds to the matrix $S$ element.
%In other words, each element $s_{ij}$ of the impact matrix corresponds to the fraction of fluid $i$ that goes to lender $j$ after this process.

We define Dependency Index ($DI_i$) of bank $i$ as a measure of the independence of a bank's portfolio of exposures in relation with the other banks portfolios:
\begin{equation}
	DI_{i} = 1- \left[\sum_{j=1}^n \left(\frac{s_{ji}}{ s_{ii} }\right)^2 \right]^{-1}.
	\label{eq:D_definition}
\end{equation}
We can also naturally define a Dependency Index for the whole system as a weighted average over $DI_i$:
\beq
	DI_{sys} = \frac{\sum_{i=1}^n\sum_{k=1}^m w_{ik} DI_i}{ \sum_{i=1}^n\sum_{k=1}^m w_{ik} }  .
	\label{eq:D_sys_definition}
\eeq

One can immediately see that $DI_i$ values are comprised in the interval $[0,1]$, and it can be directly compared with the HHI index  in (\ref{eq:HHI}), or its risk-adjusted version
\begin{equation}
	H_i = \frac{\sum_k w_{ik}^2}{\left(\sum_k w_{ik}\right)^2} .
	\label{eq:HHI_rw}
\end{equation}

It can be shown that the asymmetry of the matrix $S$ is strictly related to the heterogeneity of the total size of exposures in different institutions. 
In particular, we have that $s_{ij}\neq s_{ji}$ if and only if
\begin{equation}
	\sum_{q=1}^m w_{jq} \neq \sum_{q=1}^m w_{iq}.
\end{equation}

The two metrics, $H_i$ and $DI_i$ should be considered as complementary, as they encapsulate two different aspects of risk concentration, namely the concentration within the same portfolio and the concentration due to common exposures, respectively.

In Appendix \ref{properties-DI} we investigate the properties of the Dependency Index to check that it behaves as expected in a few simple model cases.

%\section{Applications to credit concentration and system-wide measurement}
\section{Data}

In this paper, we use several data sets to illustrate our approach.

\subsection{Data Set 1 (DS1)}
The Data Set 1 reports a sample of the real estate and corporate portfolios of two banks.
The sample covers approximatively 30\% of each banks' portfolio.
Exposures are aggregated by group, so each name corresponds to an independent counterparty.
The total number of counterparties is 1100, with only 9 of them that are common to both banks.
There are four levels of risk, associated with each counterparty, that we label from 1 (the safest) to 4 (the most risky).

\subsection{Data Set 2 (DS2)}
The network from Data Set 2 is created out of a set of leveraged loans.
We use a set of 2204 loans issued in USD and reported on a given day in the Markit database.
The data set reports several information, including the name of the issuer, the total amount issued, and the price on a given day.
Using these data, we create a set of overlapping portfolio using an approach that is described in the next Section.

\subsection{Data Set 3 (DS3)}
The Data Set 3 reports exposures from 4 banks.
%The exposures reported are all the ones that concur to the calculation of the risk-weighted assets (RWA) of each bank.
The exposures are aggregated by borrower.
More specifically, the data set reports: exposures at default (EAD), probability of default (PD), loss given default (LGD).
The data set reports two snapshots of the 4 banks in two consecutive years.

\section{Analysis of common exposures}
\label{coexp-analysis}

\subsection{Construction of the bipartite network}
The first step in constructing the bipartite network of exposures is to identify the risk-adjusted exposures.

In DS1, as we only have a discrete risk classification, we use the following method.
First, we create a homogeneous classification of counterparties' risk by matching the assessments from each bank.
Wherever the two classifications of the same counterparty do not match, we choose the highest (riskier) one.
In order to calculate the risk weighted exposures as in (\ref{eq:rw-exposures}), we consider the weight function:
\begin{equation}
	f(r) =  a + b\theta(r-r_0) ,
\end{equation}
where $a=0.2$, $b=1$ and $r_0=1.5$.
%\begin{equation}
%	f_2(r) = f_0 + \frac{1-f_0}{r_{max}-r_{min}}(r-r_{min})  ,
%\end{equation}
In practice, we consider two categories of risk ($r=1$ on one hand and $r>1$ on the other), and give to the low risk exposure a 20\% weight with respect to the high risk exposures. This mimics typical rules of calculation of risk weighted assets \cite{basel:2015bis}.

In DS2, we create a set of five  partially overlapping portfolios in the following way.
First, we choose the number of lenders (in this paper, we consider 5 lenders) and assume that each of them has a portfolio exclusively from the loans in the data set.
We aggregate the loans by issuer, calculating a risk adjusted total amount.
As a proxy of risk for each loan, we use the mid close price.
More precisely, we divide each amount by the coresponding price and consider that as a risk adjusted exposure (or sum of exposures).
Then, we randomly select 15\% of the borrowers and assign them exclusively to a randomly chosen lender.
This fraction of ``isolated borrowers'' is chosen randomly, excluding the largest 10\% loans in terms of  total amount issued.
At this point, we assign exposures using the other 85\% of issuers.
For each issuer, we divide the total amount issued into three parts, and assign each part to three randomly selected lenders.
We choose the three amount randomly, with the only condition that each tranche must not be lower than 20\% of the total amount issued.

In DS3, it is natural to use the product PD $\cdot$ EAD as risk-adjusted exposures. 

\subsection{Concentration metrics}
%\label{coexp-overlap}

\subsubsection{DS1}
The impact matrix between $A$ and $B$ is fairly symmetrical:
\beq
	S = \left(
    \begin{array}{cc}
    	0.983 & 0.019\\
    	0.017 & 0.981
    \end{array}
  \right) .
\eeq
Table \ref{tab:lender-stats-DS1} shows the calculated properties of the two lenders.
\begin{table}[htb]
	\caption{Properties of lenders of DS1. HHI and DI are the risk-weighted Herfindhal-Hirschman Index and Dependency Index, respectively. Co-exposures and co-weights are the unweighted and weighted fraction of common exposures with respect to the bank's total exposure, respectively. \label{tab:lender-stats-DS1} }
	\centering
	\begin{tabular}[t]{ccccc}
\toprule
Lender & HHI & DI & Co-exposures (\%) & Co-weights (\%)\\
\midrule
A & 0.019 & 0.00031 & 3.09 & 3.75\\
B & 0.011 & 0.00037 & 3.84 & 4.62\\
\bottomrule
\end{tabular}
\end{table}
As the fraction of portfolio overlap is only between 3 and 4\%, the impact matrix shows a weakly coupled (most of the weight is on the diagonal) and symmetric behavior.
HHI and DI move in opposite directions, as expected: bank A has a slightly more concentrated portfolio, but it is less susceptible to systemic stress, bank B has a less concentrated portfolio, but it is marginally more susceptible to the systemic risk due to common exposures.

\begin{figure}[htb]
  \centering
  \includegraphics[width=0.99\columnwidth]{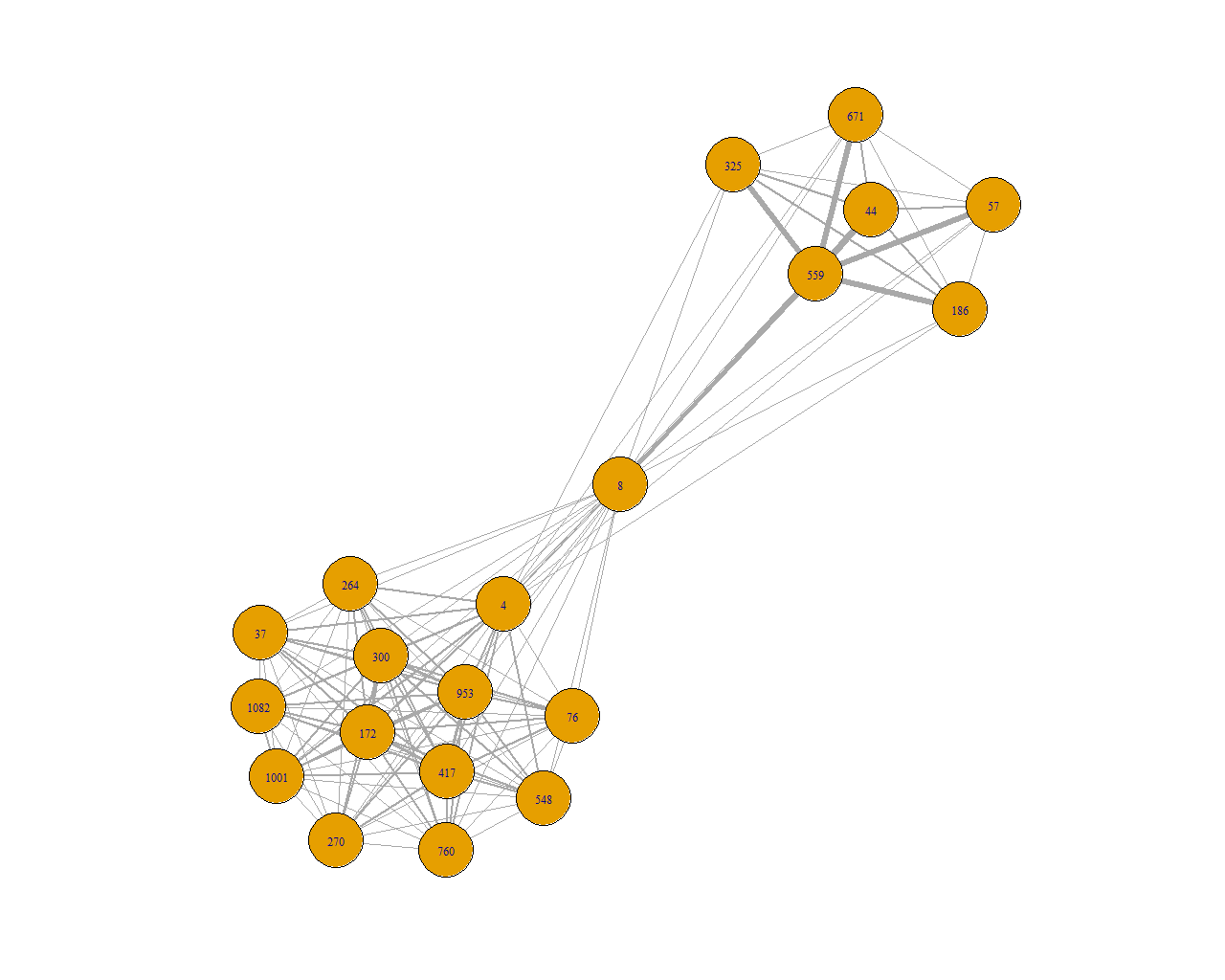}
  \caption{Projected network of interdependencies among the top 20 borrowers of banks A and B in DS1. As the two portfolios have small overlap, it is evident the clustering of the two banks portfolios. As borrowers 4 and 8 are the only ones in the overlap, they have more interconnections across the two groups.
  Borrowers that are not in the overlap tend to be more strongly connected with the borrowers in the same portfolio.}
  \label{fig:cp-network-DS1}
\end{figure}

We can investigate this further and project the bipartite network on the borrower side (Fig.~\ref{fig:cp-network-DS1} shows the network of the largest borrowers).
In this projection, the weight of the links represents the interconnectedness of two borrowers in the portfolios.
This interconnectedness increases with the size of common exposures across different lenders. 

\begin{figure}[htb]
  \centering
  \includegraphics[width=0.99\columnwidth]{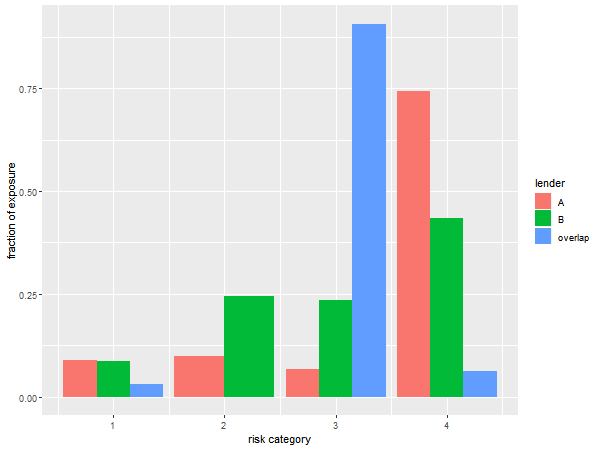}
  \caption{Risk classification of overlapping risk-adjusted exposures in DS1.}
  \label{fig:exps-per-risk-DS1}
\end{figure}

Fig.~\ref{fig:exps-per-risk-DS1} displays the composition of the portfolios in terms of risk categories.
We can see that while the risk composition of portfolios A and B are fairly similar, the overlap is much more skewed towards higher risk.
In particular, we observe that the fraction of exposures in risk category 1 (the safest) are about 9\% for bank $A$ and $B$, whereas it is only 3\% in the overlap.

\subsubsection{DS2}

In DS2, we show the impact matrix of the five lenders is:
\beq
	S = \left(
    \begin{array}{ccccc}
0.543 & 0.163 & 0.214 & 0.171 & 0.190\\
0.073 & 0.450 & 0.064 & 0.135 & 0.083\\
0.148 & 0.098 & 0.378 & 0.153 & 0.202\\
0.075 & 0.132 & 0.097 & 0.420 & 0.063\\
0.160 & 0.157 & 0.247 & 0.121 & 0.462\\
    \end{array}
  \right) .
\eeq

This synthetic data set has the highest overlap of the three, as it is also shown by the concentration properties in Table \ref{tab:lender-stats-DS2}.
\begin{table}[htb]
\caption{Properties of lenders of DS2. HHI and DI are the risk-adjusted Herfindhal-Hirschman Index and Dependency Index, respectively. The co-weights are the risk-adjusted fraction of common exposures with respect to the lenders' total exposure. 
\label{tab:lender-stats-DS2}}
\centering
\begin{tabular}[t]{cccc}
\toprule
Lender & HHI & DI &  Co-weights (\%)\\
\midrule
A & 0.127 & 0.16585 & 67.44\\
B & 0.029 & 0.27937 & 92.08\\
C & 0.101 & 0.45663 & 95.25\\
D & 0.249 & 0.29145 & 96.25\\
E & 0.015 & 0.32574 & 92.11\\
\bottomrule
\end{tabular}
\end{table}

In this system, the Dependency Index is much higher than in the other two data sets.
This is a combined effect of the high overlap (as in DS3) but also the high granularity of exposures.

\subsubsection{DS3}

The impact matrix of the four lenders is:
\beq
	S = \left(
    \begin{array}{cccc}
    	0.875 &	0.147&	0.056	& 0.307\\
		0.079 &	0.776	& 0.067	& 0.269\\
		0.003	& 0.007	& 0.863 &	0.006\\
		0.043 & 	0.070	& 0.014	& 0.419
    \end{array}
  \right) .
\eeq
As specified earlier, each off-diagonal element of the matrix $s_{ij}$ quantifies the impact of bank $i$ to bank $j$ due to the overlap among the portfolios in the system.
So, for example, bank A has a smaller impact on bank C ($s_{13} = 0.056$) than on bank D ($s_{14} = 0.307$).
On the whole, we can see that the interplay of co-exposure impacts is more complex than in DS1. 
The concentration properties of the four lenders are displayed in Table \ref{tab:lender-stats-DS3}.
\begin{table}[htb]
	\caption{Properties of lenders of DS3. HHI and D are the risk-adjusted Herfindhal-Hirschman Index and Dependency Index, respectively. Co-exposures and co-weights are the unadjusted and risk-adjusted fraction of common exposures with respect to the lenders' total exposure, respectively. \label{tab:lender-stats-DS3} }
	\centering
	\begin{tabular}[t]{ccccc}
\toprule
Lender & HHI & DI & Co-exposures (\%) & Co-weights (\%)\\
\midrule
A & 0.020 & 0.01046 & 73.34 & 75.91\\
B & 0.038 & 0.04215 & 83.58 & 66.13\\
C & 0.336 & 0.01036 & 38.22 & 44.56\\
D & 0.062 & 0.48734 & 93.67 & 97.07\\
\bottomrule
\end{tabular}
\end{table}

In this data set, there is a sizeable overlap among portfolios, however, this does not always translate into high inter-bank impacts, due to the different sizes of the portfolios involved.
For example, the impact of lender A onto lender C is much higher than the impact of lender C to lender A.
Another interesting aspect is that sometimes the HHI and DI behave differently. 
For example, lender D has a comparable HHI to lenders A and B, but a much higher interdependence, whereas lender C has a small Dependency Index due to its low interconnectedness with the rest of the system.

\subsection{Portfolio overlap and risk}
\begin{figure}
    \centering
    \begin{subfigure}{0.5\textwidth}
        \centering
        \includegraphics[width=0.9\textwidth]{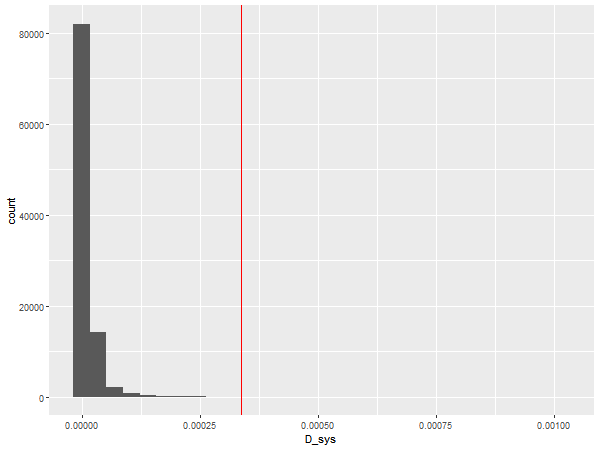}
        \caption{DS1}
        \label{fig:DsysHistDS1}
    \end{subfigure}\hfill
    \begin{subfigure}{0.5\textwidth}
        \centering
        \includegraphics[width=0.9\textwidth]{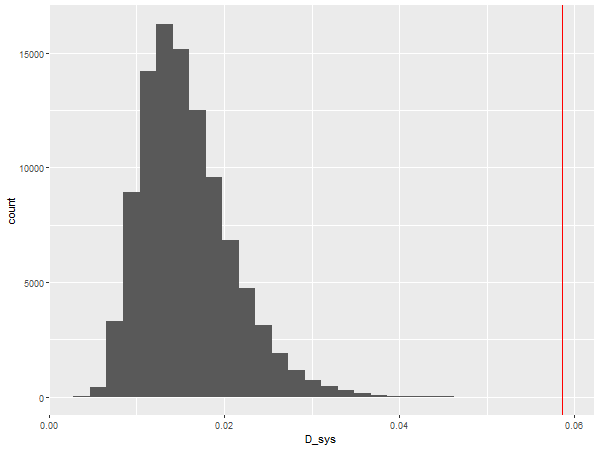}
        \caption{DS3}
        \label{fig:DsysHistDS3}
    \end{subfigure}
    \caption{Distribution of values of $DI_{sys}$ after $N=10^5$ randomizations of counterparty names within each risk category. The vertical red bar corresponds to the value calculated from the data.}
  \label{fig:Dsys-histogram}
\end{figure}

The metric of systemic interdependence $DI_{sys}$ also allows us to assess how is distributed the risk across different portfolios.
If the overlap between two portfolios were random, the risk distribution of the overlap should be similar to the one of the two portfolios.
However, Figure \ref{fig:exps-per-risk-DS1} shows that this is not the case.
Therefore, we can use the metric $DI_{sys}$ to quantify this difference.
In order to do that, we randomly rewire the network and calculate the probability of the scenario observed in the data.
We shuffle named counterparties without changing the risk composition of each lender's portfolio. 
As a result, the properties that are usually used to measure risk concentration at the lender level do not change.
At each randomization, both the HHI and the number of counterparties in the overlap remain the same, and only the link weights within risk categories are permuted.
Our metric $DI_{sys}$ of systemic interdependence, though, does change.
Fig.~\ref{fig:Dsys-histogram} shows the distribution of values of $DI_{sys}$ after $N=10^5$ randomizations for the data sets DS1 and DS3.
In both cases, the probability of observing a value larger than the value of $DI_{sys}$ measured from the data is tiny (it is $0.36\%$ for DS1 and $<0.01\%$ for DS3).
This shows that the portfolio overlap among lenders is highly non-random, and skewed towards higher interdependence.
The metric of systemic interdependence $DI_{sys}$ allows us to give a quantitative estimation of this phenomenon.

\subsection{Reaction of the system to increased credit risk}
We now test the effect that increasing the risk of counterparties may have on the portfolios interdependence in DS1.
First, we imagine a scenario where one of the two low risk (risk category 1) counterparties  in the portfolio overlap is downgraded.
Table \ref{tab:increase-risk-DS1} shows that a downgrade of either counterparty (ID 3 and 9) generates an increase of both $DI_A$ and $DI_B$.
We can study the convexity of the Dependency Index by downgrading both counterparties.
Table \ref{tab:increase-risk-DS1} also shows that the simultaneous downgrading of counterparties 3 and 9 generates an increase in $DI_A$ and $DI_B$ that is about $3.5\%$ higher than the sum of the effect of the two separate downgradings.
This shows the effect of the network in amplifying interdependence and systemic risk.
\begin{table}[htb]
	\caption{\label{tab:increase-risk-DS1} Increase of interdependence due to an increase of counterparty risk in DS1. $\Delta DI_A$ and $\Delta DI_B$ represent the difference in the Dependency Index before and after a downgrading of counterparties in the portfolio overlap. }
	\centering
	\begin{tabular}[t]{lcc}
	\toprule
	  ID & $\Delta DI_A$ & $\Delta DI_B$ \\
	\midrule
	3 & $4.7\cdot 10^{-5}$ & $5.4\cdot 10^{-5}$\\
	9 & $1.9\cdot 10^{-5}$ & $2.2\cdot 10^{-5}$\\
	Both & $6.8\cdot 10^{-5}$ & $7.9\cdot 10^{-5}$\\
	Difference & $2.4\cdot 10^{-6}$ & $2.7\cdot 10^{-6}$\\
	Difference (\%) & $3.5\%$ & $3.5\%$\\
	\bottomrule
	\end{tabular}
\end{table}

\subsection{Sensitivity of the system to an increase in overlapping borrowers}
\begin{figure}[htb]
  \centering
  \includegraphics[width=0.8\columnwidth]{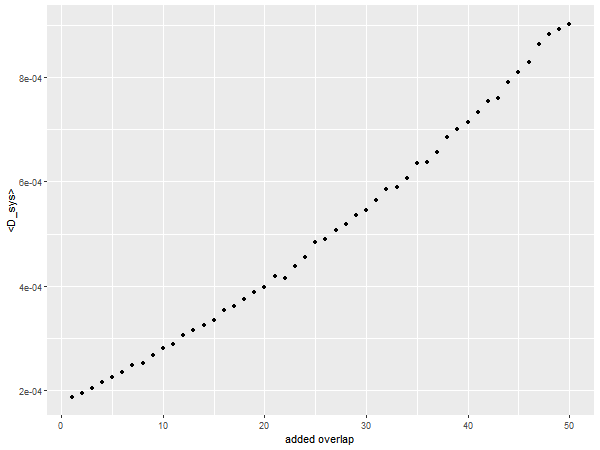}
  \caption{Evolution of $\langle DI_{sys}\rangle$ after increasing the number of overlapping borrowers. The mean is calculated over $N=1000$ iterations.}
  \label{fig:Dsys-addOverlap-DS1}
\end{figure}

We assess the effect of overlap increase by rewiring isolated exposures into common exposures in DS1.
At each iteration, we randomly choose two non-overlapping borrowers, in the same risk class, and merge them, so that they become a common exposure.
As in the previous simulations, this rewiring does not change the properties of each portfolio at the bank level, such as the HHI, the risk composition and the total exposure.
Fig.~\ref{fig:Dsys-addOverlap-DS1} shows the effect of overlap increase as measured by $DI_{sys}$.
The small non-monotonic jumps are due to the finite number of trials.
In fact, it is possible to show that the range of $DI_{sys}$ increases quite rapidly with the increase of overlap.

\subsection{Sensitivity of the system to an increase of borrowers' risk}

A way to characterize the systemic importance of single borrowers is to assess the effect that an increase in their risk may have on global metrics.
In order to do that, for each borrower  we calculate the difference in $DI_{sys}$ and $HHI_{sys}$ after having increased by a factor 5 all the exposures corresponding to the considered borrower.
In this way, we estimate the effect that an increase in its risk has on the concentration parameters of the system.
The choice of the factor 5 mimics a typical risk weight prescription of giving 20\% of EAD to top tier counterparties and 100\% to other counterparties.

\begin{figure}
    \centering
    \begin{subfigure}{0.45\textwidth}
        \centering
        \includegraphics[width=0.99\textwidth]{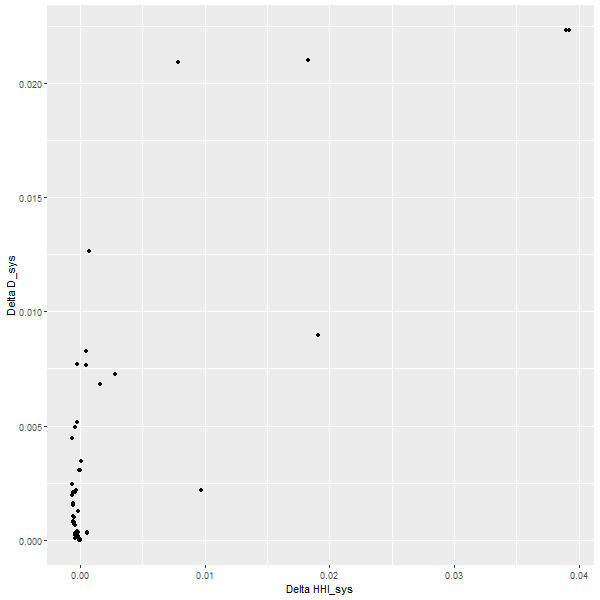}
        \caption{DS3, year 1}
    \end{subfigure}\hfill
    \begin{subfigure}{0.45\textwidth}
        \centering
        \includegraphics[width=0.99\textwidth]{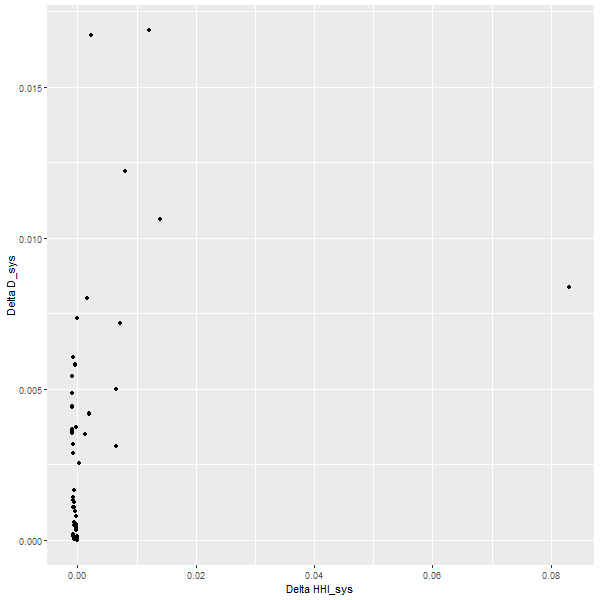} 
        \caption{DS3, year 2}
    \end{subfigure}
    \caption{Sensitivity of the system to an increase in borrowers risk in DS3. Each point represents a borrower whose risk has been increased by multiplying their adjacent exposures by a factor 5.  }
  \label{fig:deltaDsys-deltaHHI-borrRisk-DS3}
\end{figure}

Figure \ref{fig:deltaDsys-deltaHHI-borrRisk-DS3} shows the result of this test in DS3.
We plot the set of points with $\Delta DI_{sys} > 0$, corresponding to the borrowers in the overlap.
From this scatterplot it transpires that a narrow interval in $\Delta HHI_{sys}$ may correspond to a quite diverse behavior in terms of $\Delta DI_{sys}$.
This is another sign of the extra information brought in by the new metric.
Besides, the different values of $\Delta DI_{sys}$ provide information on which borrowers are more systemically important whenever their risk should increase.
We will use this information to calculate the common exposure add-on in Section \ref{additional-capital}.

We have also compared the systemic interdependence sensitivity  $\Delta DI_{sys}$ with well known network centrality measures, to verify if the information captured by $\Delta DI_{sys}$ were encapsulated in more common network measures as in \cite{Das:2016ii}.
While eigenvector centrality appears to resemble the behavior of $\Delta DI_{sys}$ in one data set, we do not find any conclusive evidence that this holds across all the data sets we consider.

\section{Additional capital requirements of common exposures}
\label{additional-capital}

The Internal Ratings Based (IRB) capital requirement and the Granularity Adjustment (GA) formula provide an estimation of the capital needed by an institution to be functioning in the occurrence of economic downturns.
In Appendix \ref{IRB-GA-summary} we give a summary of the relevant formulae of regulatory capital and Granularity adjustment that we are using in this paper.
In this section, we wish to calculate a third term, the further additional capital that is required to consider the systemic risk due to common exposures.

\subsection{Common exposure adjustment}

Let us consider a lender with a portfolio of $N$ borrowers in a system of lenders with overlapping portfolios.
In order to estimate the co-exposure capital add-on, we need a quantity that can give us a measure of the relevance of each borrower in the common exposure risk.
As we have seen in Section \ref{coexp-analysis}, the increment in the Dependency Index under credit risk increase ($\Delta DI_i$) measures exactly that.
In analogy with (\ref{eq:IRB-Ki}), we combine the PD and the Dependency Index and define the corresponding capital requirement as 
\beq
	\label{eq:XCE}
	X_{CE} = \sum_{i=1}^N s_i X_{CE}^i,
\eeq
where $s_i$ is the fraction of exposures of a given lender to borrower $i$ (i.e.~$\sum_i s_i=1$)
and
\beq
	X_{CE}^i = MA_i \cdot LGD_i \left\{ \Phi\left[ \Phi^{-1}(PD_i) +\eta\;\Phi^{-1}\left(\frac{1+\theta(\Delta DI_i) \Delta DI_i}{2}\right)  \right] - PD_i \right\},
\eeq
where $\theta(x)$ is the step function; in other words, we only consider positive increments of Dependency ($\Delta DI_i>0$). When a borrower $i$ is not in the overlap, i.e. only linked to a single institution, its contribution to $X_{CE}$ is zero (note that $\Phi^{-1}(0.5) = 0$).
This formula depends on the parameter $\eta$, that governs the weight of the systemic co-exposure effect with respect to the other capital requirements.

The measure in (\ref{eq:XCE}), however, needs to be calibrated to avoid that part of the risk encapsulated in this term is already considered in the GA term.

\subsection{Example of double counting}

%This typically occurs when the Granularity Adjustment increases as a function of the overlap.
%If this happens for a particular system, it means that the GA is able to capture, at least in part, the concept that larger overlap among portfolios requires more additional capital.
%On the other hand, for many systems, it is quite possible that also the opposite way may be true: the GA \emph{decreasing} as  the overlap increases.
%This can typically occur when the overlap is small enough that a small increase actually improves the granularity of the single portfolios, for example if exposures are transfered from large to small counterparties.

\begin{figure}[htb]
    \centering
    \begin{subfigure}[b]{0.8\textwidth}
        \centering
        \includegraphics[width=0.99\columnwidth]{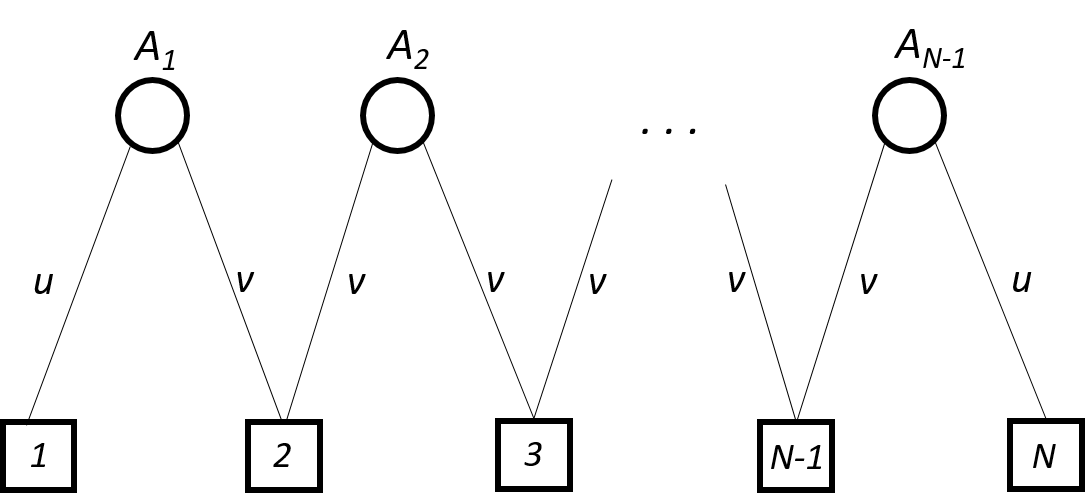}
	        \caption{Example of an uniformly co-exposed system of lenders.}
        \label{fig:uvw-diagram-A}
    \end{subfigure}%
    \\ 
    \begin{subfigure}[b]{0.8\textwidth}
        \centering
        \includegraphics[width=0.99\columnwidth]{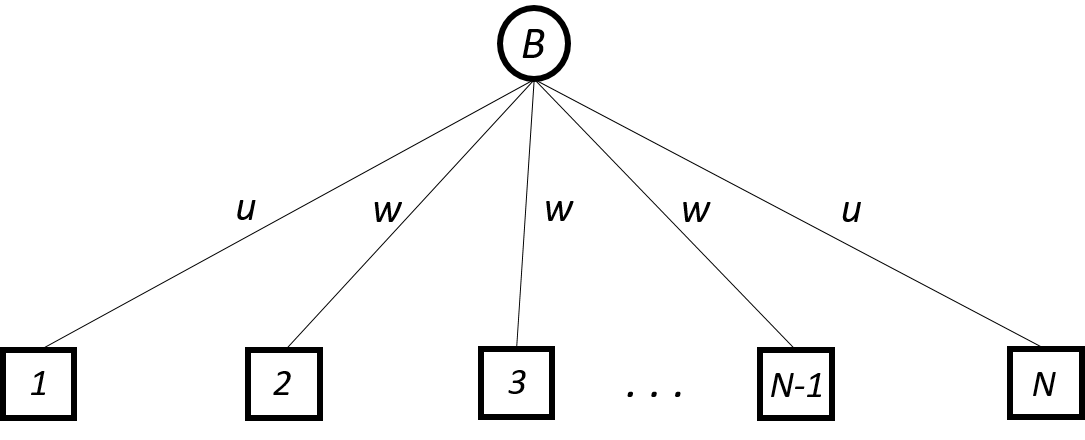}
        \caption{Corresponding system where all the lenders have been aggregated into a ``superlender'' B.}
        \label{fig:uvw-diagram-B}
    \end{subfigure}
    \caption{Example of a comparison between a simple system of overlapping portfolios (a) and a corresponding aggregated system (b), with one ``superlender'' that has the same counterparties and the same aggregated exposures ($w=2v$).}
\end{figure}

Let us consider the particular case represented in Figure \ref{fig:uvw-diagram-A}. 
$(N-1)$ lenders are exposed to $N$ borrowers; each lender has two counterparties; each exposure has the same weight $v$ except the two isolated exposures that have each weight equals to $u$.
In order to compare the Dependency Index and the Granularity Adjustment, we define a corresponding system where we aggregate all the lenders into one and sum all the corresponding exposures.
The result is shown in Figure \ref{fig:uvw-diagram-B}.
The common exposures become exposures of the aggregated lender and each has weight $w=2v$, whereas the two isolated exposures remain the same ($u$).
As we assume normalized weights, we have $2u+(N-2)w = 1$.
Due to the homogeneity of common exposures in system (a), we can calculate the GA in system (b) and study its behaviour in terms of the parameters $u$ and $w$.
By construction, the parameter $w=2v$ corresponds to the overlap of portfolios in system (a), and the parameter $u$ corresponds to the non overlapping part.

\begin{figure}[htb]
  \centering
  \includegraphics[width=0.5\columnwidth]{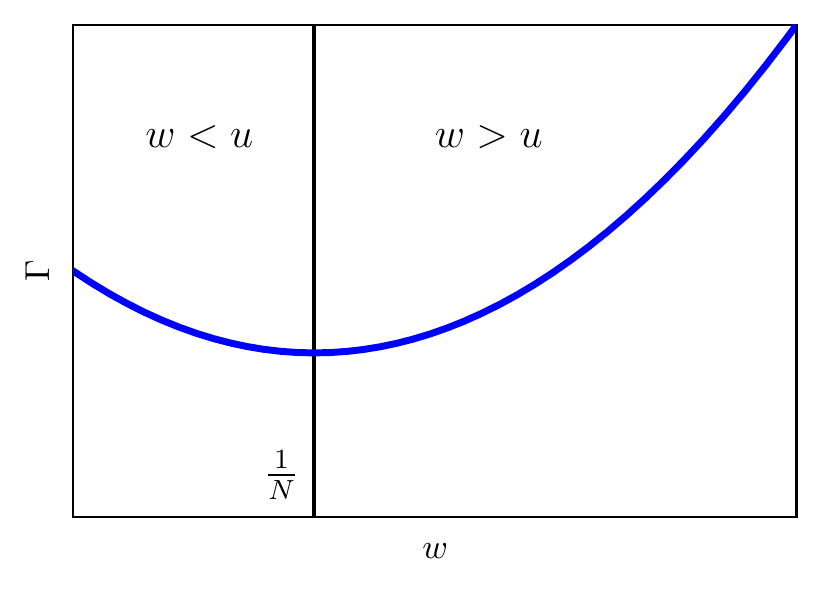}
  \caption{Graphical representation of Equation (\ref{eq:example-gamma}).}
  \label{fig:plot-gamma-w}
\end{figure}

A simple calculation of the Granularity Adjustment $\Gamma$ of system (b) yields:
\beq
\label{eq:example-gamma}
	\Gamma = constant \left[ 1 - 2(N-2)w + N(N-2)w^2 \right].
\eeq
The behaviour of Equation (\ref{eq:example-gamma}) is represented in Figure \ref{fig:plot-gamma-w}.
In this particular example, the behaviour of $\Gamma(w)$ corresponds to the behaviour of the GA as a function of the overlap in system (a). 
It  is easy to see that there are two regimes. 
When $w>u$, the GA encapsulates the effect of the overlap, because an increase in the overlap and therefore the interdependence of system (a) (i.e. an increase in $w$)  corresponds to an increase in the concentration of the portfolio of lender $B$ in system (b).
When $w<u$, the GA and the common exposures move in opposite directions, because an increase in the overlap (i.e. an increase in $w$) corresponds to a more homogeneous portfolio for lender $B$.
In other words, we can infer that the condition of double counting is 
\beq
\label{eq:inferred-double-counting}
	\frac{\partial\Gamma}{\partial(overlap)} > 0.
\eeq

\subsection{Derivation of the capital adjustment}

We use then this observation and equation (\ref{eq:inferred-double-counting}) to estimate the double counting between the GA and the Dependency Index.
Let us distinguish whether the relative exposures $s_i$ are in the overlap (i.e. they are shared by at least two lenders) or not, by defining $\Omega$ the set of borrowers in the overlap and $Z$ the set of borrowers not in the overlap, and re-write the Granularity Adjustment (\ref{eq:GA-formula}) as
\beq
\label{eq:rewritten-GA-formula}
	\Gamma =  \frac{\sum_{i\in \Omega} s_i^2 \tilde{K}_i +  \sum_{i\in Z} s_i^2 \tilde{K}_i }
	{\sum_{i\in \Omega} s_i {K}_i +  \sum_{i\in Z} s_i {K}_i},
\eeq
where $K_i$ is the IRB approach capital corresponding to borrower $i$, and we define
\beq
	 \tilde{K}_i = \frac{1}{2} C_i \left[ (\delta-1)K_i + R_i \right].
\eeq

Let us know define the weighted means of $K_i$ over each subset $\Omega$ and $Z$ as
\beq
	 \langle K \rangle_\Omega = \frac{\sum_{i\in \Omega} s_i {K}_i}{\sum_{i\in \Omega} s_i}  \qquad \langle K \rangle_Z = \frac{\sum_{i\in Z} s_i {K}_i}{\sum_{i\in Z} s_i} .
\eeq
Using these quantities, we can approximate the formula (\ref{eq:rewritten-GA-formula}) as
\beq
%\label{eq:}
	\Gamma =  \frac{\left( \sum_{i\in \Omega} s_i^2 \right) \langle \tilde{K} \rangle_\Omega + \left( \sum_{i\in Z} s_i^2 \right) \langle \tilde{K} \rangle_{Z} }
	{ \left( \sum_{i\in \Omega} s_i \right) \langle K \rangle_\Omega + \left( \sum_{i\in Z} s_i \right) \langle K \rangle_{Z}}.
\eeq

In order to study how the GA behaves with respect to overlap, we assume a small perturbation $ \bf{\epsilon}$ of the exposures fractions $\bf{s}$: $\Gamma(\bf{s}^{\prime}) =	\Gamma (\bf{s + \epsilon})$.
The perturbation slightly increases  the overlap and decreases the non-overlap by the same amount, more specifically
\beq
\label{eq:perturbation}
	\begin{cases} 
		s_i^{\prime} = s_i + \epsilon & \mbox{if } i\in\Omega \\ 
		s_i^{\prime} = s_i - \frac{N_{\Omega}}{N_Z} \epsilon & \mbox{if } i\in Z
	\end{cases}
\eeq
where $\epsilon>0$,  $N_{\Omega}$ is the number of borrowers in the overlap and $N_Z = N-N_{\Omega}$ is the number of borrowers outside the overlap.
This perturbation increases overlap without changing the total exposures of the considered lender ($\sum_i s_i= \sum_i s_i^{\prime} = 1$).

After imposing the conditions (\ref{eq:perturbation}), the overlap in the GA is encapsulated by the scalar parameter $\epsilon$ and we can study how the GA increases with it:
\beq
	\label{eq:GA-expansion}
	\left. \frac{\mbox{d}\Gamma}{\mbox{d} \epsilon}\right|_{\epsilon=0}  \geq 0.
\eeq

Equation (\ref{eq:GA-expansion}) is satisfied  if and only if
\beq
	\label{eq:r-doublecounting}
	r = \frac{N_{\Omega}}{N_Z} 
	\frac{
			2 \langle \tilde{K} \rangle_{Z} \sum_{i \in Z} s_i \left( \langle K \rangle_{\Omega} \sum_{i\in\Omega} s_i + \langle K \rangle_Z \sum_{i\in Z}s_i  \right) + 
			N_Z \langle K \rangle_{\Omega} \left( \langle \tilde{K} \rangle_{\Omega} \sum_{i\in\Omega} s_i^2 + \langle \tilde{K} \rangle_Z \sum_{i\in Z}s_i^2  \right) 
		}
	{
			2 \langle \tilde{K} \rangle_{\Omega} \sum_{i \in Z} s_i \left( \langle K \rangle_{\Omega} \sum_{i\in\Omega} s_i + \langle K \rangle_Z \sum_{i\in Z}s_i  \right) + 
			N_{\Omega} \langle K \rangle_{Z} \left( \langle \tilde{K} \rangle_{\Omega} \sum_{i\in\Omega} s_i^2 + \langle \tilde{K} \rangle_Z \sum_{i\in Z}s_i^2  \right) 
		} 
	\leq 1.
\eeq
%where
%\beq
%\eeq

 Equation (\ref{eq:r-doublecounting}) gives an  estimate on the amount of double counting between the GA and our co-exposure capital $X_{CE}$, to correct the capital adjustment.
We propose then a total capital requirement that reads
\beq
	K_{total} = K + \Gamma + K_{CE}(\alpha,\eta),
\eeq
where the Common Exposure Capital Adjustment $K_{CE}$ is
\beq
	\label{eq:KCE}
	K_{CE}(\alpha,\eta) = [\alpha(r-1) + 1] X_{CE}(\eta).
\eeq
The two parameters ($\alpha$ and $\eta$)  govern the capital add-on due to common exposures.
The term $[\alpha(r-1) + 1]$ in (\ref{eq:KCE}) is designed in such a way that when double counting occurs ($r<1$), the co-exposure capital requirement decreases linearly with $r$ (and therefore $K_{CE} < X_{CE}$).

\section{Parameter calibration and application to data}

The calibration of parameters $\alpha$ and $\eta$ is obtained by comparison with numerical simulations.

\subsection{Numerical simulations}
We perform numerical simulations using as an input the PDs reported in DS3, the data set with the largest overlap.
For each lender, we calculate the distribution of losses by uniformly generating random  numbers for each borrower $i$ and comparing it with their corresponding $PD_i$.
In the present work we use $10^5$ iterations.
We then calculate the Value-at-Risk (VaR) at $q=0.999$ and the expected losses EL.
The unexpected losses are simply the difference $UL = VaR - EL$.

We perform these simulations both with the reported data and in a simplified downturn scenario.
The downturn scenario is encapsulated by a new set of probabilities of default $PD^{S}$ that we define as
\beq
	PD_i^S = \sqrt{A \cdot PD_i},
\eeq
where $A = 0.3$.
We adopt this definition to mimic the effect of a downturn to probabilities of default.

%[show distributions?]

\subsection{Parameter calibration}

In order to calibrate the parameters $\alpha$ and $\eta$, we choose a data set and compare the results of the numerical simulations in the downturn scenario with the analytical capital requirements resulting from the sum of the capital calculated in the IRB infinitely granular approach (\ref{eq:basel}) plus the Granularity Adjustment (\ref{eq:GA-formula}).
The difference between the analytically calculated  capital and the estimation from the numerical simulations constitutes a gap that we use to calibrate the parameters, by setting this gap equal to the common exposure adjustment $K_{CE}$.

This method is not driven by the quite strong assumption that the gap can be entirely explained by the systemic effect of common exposures.
Rather, it aims to use the gap as a guide towards a typical scale to which it seems sensible to measure a further capital requirement.
In real terms, the systemic effect of  common exposures may be larger than what can be captured by Monte Carlo simulations that, differently from the dependency index and analogous network measures, focus on one portfolio at the time.

Table \ref{tab:compare-MC-formula} calculates this gap using Data Set 3 (year 1).
\begin{table}[htb]
	\caption{Calculation of the difference (gap) between the unexpected losses (UL) calculated by Monte Carlo simulations in the adverse scenario described in the text and the analytical calculation of regulatory capital ($K$) plus granularity adjustment ($\Gamma$), using Data Set 3 (year 1). \label{tab:compare-MC-formula} 
	}
	\centering
	\begin{tabular}[t]{cccc}
	\toprule
	Lender & UL (numeric) & $K+\Gamma$ & Gap \\
	\midrule
	A & 0.043 & 0.036 & 0.007\\
	B & 0.051 & 0.043 & 0.008\\
	C & 0.232 & 0.372 & -0.139\\
	D & 0.074 & 0.065 & 0.009\\
	\bottomrule
	\end{tabular}
\end{table}

With the exception of lender $C$, the analytically calculated capital requirement ($K+\Gamma$) is smaller than the unexpected losses simulated numerically.
In the case of lender $C$, the level of risk of the portfolio appears instead to be completely captured by $K+\Gamma$. Therefore, in the case of lender $C$ we set $K_{CE}=0$. 
For the other three lenders, instead, we set the gap equal to the co-exposure capital $K_{CE}$ in (\ref{eq:KCE}) and use the method of the least squares to fit the parameters.
We obtain:
\beq
	\label{eq:fitted-parameters}
	\alpha = 0.53; \qquad \eta = 68.9.
%	\alpha = 0.62; \qquad \eta = 58.6.
\eeq
Using the parameters in (\ref{eq:fitted-parameters}) we can then calculate $K_{CE}$ for the other year of DS3.
Table \ref{tab:calculated-KCE} shows the results of this calculation as well as a comparison with the numerical calculation.
\begin{table}[htb]
	\caption{Calculation of the common exposure capital addition $K_{CE}$ for the system reported in DS3 (year 2).
	\label{tab:calculated-KCE}
	}
	\centering
	\begin{tabular}[t]{cccc}
	\toprule
	Lender & UL (numeric) & $K+\Gamma$ & $K_{CE}$ \\
	\midrule
	A & 0.0732 & 0.0607 & 0.0136\\
	B & 0.0392 & 0.0339 & 0.0039\\
	C & 0.1312 & 0.1741 & 0.0000\\
	D & 0.0714 & 0.0647 & 0.0045\\
	\bottomrule
	\end{tabular}
\end{table}
As we can see, after calibration with a different data set, the capital add-on $K_{CE}$ is able to capture quite well the  amount of difference between $K+\Gamma$ and the simulated UL.

\section{Conclusions}
\label{conclusions}

In this paper, we propose a new measure of credit concentration risk (Dependency Index) that encapsulates the systemic effect of interconnectedness due to common exposures. 
We apply this metric on several data sets to illustrate its properties, and we use the metric to calculate analytically the additional capital that quantifies the systemic effect of common exposures risk.
%This additional capital is additional to both the original capital requirement from Basel II and the granularity adjustment later developed.
The capital add-on can be calibrated to avoid double counting between the granularity adjustment and the co-exposure adjustment.
%The methodology to avoid double counting is quite general and it can be used in other capital adjustments, as it only depends on general properties of the  analytical description of the capital requirement.

Broadly speaking, our approach is to measure credit concentration risk at the system level, rather than separately looking at  each financial institution portfolios individually.
We show that this metric  is able to summarize the degree of interdependence in a network of exposures.

The value of the Dependency Index is essentially affected by the network topology and the riskiness of exposures.
By testing  this metric  on a number of data sets, even in the case of low overlap among portfolios, we show that the metric is able to describe the effect of credit risk increase  and capture non-linear effects due to the interplay of exposure connectivity.
The interdependence analysis also provides insights into the properties of the  overlap itself.
In particular, we show that  the portfolio overlap between  two institutions may be highly non-random, and skewed towards higher interdependence.
This may be due to a number of reasons: tendency of high risk borrowers to be indebted to multiple institutions, lower credit rating of larger borrowers, etc.
Investigating these causal relationships is beyond the scope of the present work, but the Dependency Index allows us to give a quantitative estimation of this phenomenon.
We also explore the effect of overlap increase and find that the metric behaves as expected.
By comparing the sensitivity of DI and HHI to an increase of borrowers' risk, we show that the Dependency Index provides extra information on which borrowers are systemically important.

Although approximated, our co-exposure adjustment is able to capture, with only two parameters, an aspect of systemic risk that, to our knowledge, has been neglected so far and the complexity in the interplay among (risk-adjusted) exposures.
This is an essential step in developing a more comprehensive methodology to supervise and manage credit risk in a complex system of financial institutions.

\section*{Acknowledgements}
We thank  Michael Gordy, Eugene Kashdan, Pauric McBrien, Leo Regan and Adrian Varley, for useful discussions and feedback.
Any remaining errors are of the authors.

%%%%%%%%%%%%%%%%%%%%%%%%%%%%%%%%%%%%%%%%%%%%%%%%%%%%%%%%%%%%%

\begin{appendices}

\section{Properties of the Dependency Index}
\label{properties-DI}

Metrics of inequality are often tested against a number of properties \cite{calabrese:2012ucd,lorenz:1905asa,lutkebohmert:2008s}. Therefore, here we analyse some properties of the Dependency Index $DI$ to check its validity as a measure of co-exposure risk.

\subsection{Minimum Dependency}
\emph{Dependency $DI_i$ of lender $i$  is zero if and only if when lender $i$ has no co-exposures.}

We can define this concept more formally as
\beq
	DI_i=0  \qquad \Leftrightarrow  \qquad  s_{ji}=0 \qquad \textrm{for each} \qquad j\neq i.
\eeq
This is immediately evident from the definition (\ref{eq:D_definition}).

\subsection{Transfer to co-exposures}
\label{transfer}
\emph{If a lender transfers a fraction of an exposure from an isolated borrower to a co-exposed borrower, by an amount that is small compared with the exposure size, the dependency of that lender increases.}

\begin{figure}
  \centering
  \includegraphics[width=0.6\columnwidth]{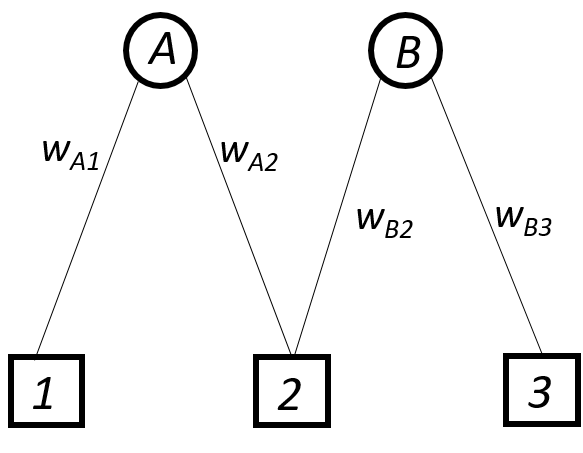}
  \caption{Building block of a bipartite network of common exposures.  }
  \label{fig:simplest-bipartite}
\end{figure}

We can see this by focusing on a bipartite network with two lenders ($A$ and $B$), three borrowers (1,2,3) and the following matrix of exposures:
\begin{displaymath}
  W =
  \left(
    \begin{array}{cc}
    	w_{A1} & 0\\
    	w_{A2} & w_{B2}\\
    	0 & w_{B3}
    \end{array}
  \right) .
\end{displaymath}
This is the building block of a bipartite network of common exposures (Fig.~\ref{fig:simplest-bipartite}).

If we transfer a quantity $\epsilon>0$ from the isolated exposure $1$ to the common exposure $2$ as in
\begin{displaymath}
  \left\{
    \begin{array}{ccc}
    	w'_{A1} &=& w_{A1}  - \epsilon\\
    	w'_{A2} &=& w_{A2}  + \epsilon
    \end{array}
  \right. ,
\end{displaymath}
we can study the effect on the dependency indexes as a function of $\epsilon$.
A direct calculation shows that $\partial D_i(\epsilon)/\partial\epsilon \geq 0 $ for $i=A, B$, so does $DI_{sys}$ by definition.

\subsection{Merging of common exposures (superadditivity)}
\emph{If two common exposures merge, the dependency metric of the involved lenders increases.}

\begin{figure}
    \centering
    \begin{subfigure}{0.45\textwidth}
        \centering
        \includegraphics[width=0.9\textwidth]{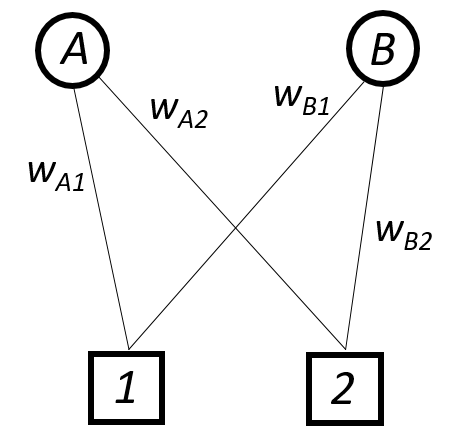}
        \caption{Before the merge}
        \label{fig:before-merge}
    \end{subfigure}\hfill
    \begin{subfigure}{0.45\textwidth}
        \centering
        \includegraphics[width=0.8\textwidth]{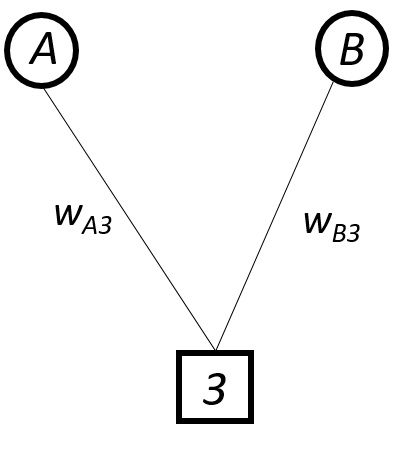} 
        \caption{After the merge}
        \label{fig:after-merge}
    \end{subfigure}
    \caption{Merging of two common exposures. Borrowers 1 and 2 merge and become borrower 3, with $w_{A3} = w_{A1}+w_{A2} $, $w_{B3} = w_{B1}+w_{B2} $.  }
  \label{fig:bipartite-two-coexp}
\end{figure}
Let us consider the simplified case of a network with two lenders and two common exposures (Fig.~\ref{fig:before-merge}), described by the weight matrix:
\begin{displaymath}
  W =
  \left(
    \begin{array}{cc}
    	w_{A1} & w_{B1}\\
    	w_{A2} & w_{B2}
    \end{array}
  \right) .
\end{displaymath}
We wish to study the difference in the dependency metrics with the corresponding network characterized by a merging of the two borowers, {\ie}
\begin{displaymath}
  W' =
  \left(
    \begin{array}{cc}
    	w_{A3} & w_{B3}
    \end{array}
  \right) ,
\end{displaymath}
where $w_{A3} = w_{A1}+w_{A2} $, $w_{B3} = w_{B1}+w_{B2} $ (Fig.~\ref{fig:after-merge}).

A direct calculation shows that we have always $D'_A \geq D_A$ and $D'_B\geq D_B$.
In particular, we have $D'_A = D_A$ and $D'_B = D_B$ if and only if $w_{A1}/w_{A2} = w_{B1}/w_{B2}$, {\ie} the merging of two common exposures typically increases the systemic interdependence. The exception is a  particular case when the relative exposure of the two lenders to the two borrowers is exactly the same.

\subsection{Effect of isolated exposures}
\emph{If a lender transfers a fraction of an exposure to a new counterparty that has not borrowed from other lenders, by an amount that is small compared with the exposure size, the Dependency Index of the system decreases.}

To see that, we can use the same setting as in Section \ref{transfer}, where $w_{A1}=0$ and we consider the transfer
\begin{displaymath}
  \left\{
    \begin{array}{ccl}
    	w'_{A1} &=& \epsilon\\
    	w'_{A2} &=& w_{A2}  - \epsilon
    \end{array}
  \right. ,
\end{displaymath}
where $\epsilon>0$.
A direct calculation shows that $D'_A \leq D_A$, $D'_B\leq D_B$, and therefore $D'_{sys} \leq DI_{sys}$ by definition.

\section{IRB approach and Granularity Adjustment}
\label{IRB-GA-summary}

Let us summarize here the Internal Ratings Based (IRB) approach and the formula of Granularity Adjustment that we use in the paper.
Our starting point is recalling the formula of  capital calculation according to the IRB approach Basel methodology \cite{basel:2017bis}:
\beq
	\label{eq:basel}
	K = \sum_{i=1}^N s_i K_i,
\eeq
where the sum runs over all the exposures of the considered lender, $s_i$ is the fraction of exposure to borrower $i$, and $K_i$ is defined as
%\footnote{From now on we assume an effective maturity of one year, so to avoid a maturity adjustment coefficient  }
\beq
	\label{eq:IRB-Ki}
	K_i = MA_i \left\{LGD_i\cdot \Phi\left[ \frac{\Phi^{-1}(PD_i) + \sqrt{\rho_i}\Phi^{-1}(q) }{ \sqrt{1-\rho_i}}  \right] - LGD_i  \cdot PD_i \right\},
\eeq
where $\Phi(x)$ is the cumulative normal distribution and $\Phi^{-1}(x)$ its inverse, $q$ is the $q^{th}$ chosen percentile of the (normally distributed) systematic risk factor, $\rho_i$ is the correlation between the returns of borrower $i$ and the systematic risk factor, and $MA_i$ is the maturity adjustment, that is estimated as 
\beq
	MA_i= \frac{1+(M-2.5)b(PD_i)}{1-1.5b(PD_i)},
\eeq
where
\beq
	b(PD) = \left[ 0.119 - 0.0548 \log(PD) \right].
\eeq
Typically, one sets $q=0.999$.

As it is well-known, the Basel formula (\ref{eq:basel}) is written for a portfolio that is infinitely granular.  
It is then necessary to add a Granularity Adjustment (GA).
Here we use the Granularity Adjustment  proposed by \cite{Gordy:2013}:
\beq
\label{eq:GA-formula}
	\Gamma = \frac{1}{2K} \sum_{i=1}^N s_i^2 C_i \left[ \delta(K_i + R_i) -K_i \right],
\eeq
where $R_i$ is the loan-loss reserve
\beq
	R_i = E[LGD_i] \cdot PD_i,
\eeq
$\delta$ is a regulatory parameter that is calibrated as $\delta=4.83$
and $C_i$ is defined as
\beq
	C_i = \frac{\gamma E[LGD_i](1-E[LGD_i]) + E[LGD_i]^2}{ E[LGD_i]},
%	C_i = \frac{\gamma LGD_i(1-LGD_i) + LGD_i^2}{ LGD_i},
\eeq 
where $\gamma=0.25$ \cite{Gordy:2013}.

\end{appendices}

%----------------------------------------------------------------------------------------
%	BIBLIOGRAPHY
%----------------------------------------------------------------------------------------
%\clearpage
%\printbibliography[title={Bibliography}] % Print the bibliography, section title in curly brackets

%----------------------------------------------------------------------------------------

\clearpage
%\bibliographystyle{elsarticle-harv}\biboptions{authoryear}
%\section*{Bibliography}
\bibliographystyle{elsarticle-harv}
\bibliography{NetBib2}

\end{document}